\documentclass[prl,twocolumn,preprintnumbers,amsmath,amssymb]{revtex4}

\usepackage{latexsym}
\usepackage{hyperref}

\def\p{\partial}
\def \ra{{\rightarrow}}
\newcommand{\nuo}{{r}_++{r}_--{1\over\nu}\sqrt{{r}_+{r}_-(\nu^2+3)}}

\newcommand{\bea}{\begin{eqnarray}}
\newcommand{\eea}{\end{eqnarray}}
\newcommand{\be}{\begin{equation}}
\newcommand{\ee}{\end{equation}}

\newcommand{\SL}{\mbox{SL}(2,\mathbb{R})}
\newcommand{\re}[1]{(\ref{#1})}

\def\eps{\epsilon}

\def\d{\partial}

\def\cK{{\cal K}}
\def\cL{{\cal L}}

\def\cQ{{\cal Q}}

\def \eps{{\epsilon}}

\begin{document}
\title{Inner Mechanics of 3d Black Holes}

\author{St\'ephane Detournay}


\affiliation{%
Center for the Fundamental Laws of Nature, Harvard University,\\
Cambridge, MA 02138, USA}

%


\begin{abstract}

We investigate properties of the inner horizons of certain black holes in higher-derivative three-dimensional gravity theories. We focus on BTZ and Spacelike Warped Anti-de Sitter black holes, as well as on asymptotically Warped de-Sitter solutions exhibiting both a cosmological and a black hole horizon. We verify that a First Law is satisfied at the Inner horizon, in agreement with the proposal of \cite{Castro:2012av}. We then show that, in Topologically Massive Gravity, the product of the areas of the inner and outer horizons fails to be independent on the mass, and trace this to the diffeomorphism anomaly of the theory.

\end{abstract}


\maketitle

\begin{center}
\section{Introduction and Summary}
\end{center} 
The derivation of the laws of black hole mechanics and the observation that their horizon area behaves like an entropy \cite{Bekenstein:1972tm,Bardeen:1973gs,Hawking:1974sw} have both played a central role in recognizing the thermodynamic nature of black holes. 
A major challenge consists in going beyond the thermodynamic regime provided by classical (super)gravity theories and identify the microscopic degrees of freedom responsible for their enormous entropy, possibly paving the way towards a quantum theory of gravity. Many fruitful steps have been taking towards this objective --with or without string theory-- starting with \cite{Strominger:1996sh, Strominger:1997eq,Sen:1995in}, where the derivation essentially relied on the presence of an $AdS_3$ factor and on the application of the Cardy formula to its associated two-dimensional conformal symmetry \cite{Brown:1986nw}. It is now believed that the derivation is much more general and should apply to a wider class of black hole solutions, including astrophysically relevant ones \cite{Guica:2008mu}.

While the geometric quantity associated with black hole entropy seems naturally to be the outer horizon, there have been indications towards the relevance of black holes {\itshape inner} horizons. The first hint comes from the fact that for any asymptotically flat black hole admitting a smooth extremal limit, the product $A_+ A_-$ of the inner and outer horizons' areas seems to depend only on the quantized charges and is independent of the mass \cite{Larsen:1997ge} (see also \cite{Castro:2009jf,Cvetic:2010mn, Ansorg:2008bv, Ansorg:2009yi,Castro:2012av}). Another intriguing feature put forwards recently \cite{Castro:2012av} (see \cite{Curir1, Curir2} for related observations in the Kerr black hole) is the fact that inner horizons seem to enjoy their own first law, in the form
\be
  -dM  = T_- dS_- - \Omega_- dJ + \cdots
\ee 
where the intensive quantitites are computed at the inner horizon. While the implications of this relation are still unclear, it has been checked for a fair amount of  examples for Einstein gravity coupled to matter fields, for asymptotically flat black objects in 5 dimensions, including black strings and black rings \cite{Castro:2012av}. One might wonder if a similar relation holds for higher derivative gravity theories, for which
the First Law for the outer horizon holds in all known situations, and for black holes with different asymptotics. 
We investigate the properties of inner horizons of various black holes in a simple higher derivative theory in three dimensions, Topologically Massive Gravity (TMG) \cite{Deser:1981wh}. We consider BTZ, Warped $AdS_3$ and Warped $dS_3$ black holes, showing that the observations of \cite{Castro:2012av} naturally extend in this context. The latter case demonstrates that the reasoning is also applicable to solutions with both a cosmological and a black hole horizon. We compute the product of the inner and outer entropies and observe that the product does depend on the mass, and is therefore no longer quantized. We note that the mass dependance originates from the diffeomeophisms anomaly of TMG, proportional to the differences of the central charges \cite{Kraus:2005zm}. We comment on the extension of these results to generic three-dimensional higher-derivative theories, in particular New Massive Gravity \cite{Bergshoeff:2009hq}.
%
%
\newline
\newline
\section{Warm-up: BTZ black holes in pure GR}
Consider the BTZ metric \cite{Banados:1992wn, Banados:1992gq}
\begin{equation}\label{BTZ}
 ds^2_{BTZ} = -(N^\perp)^2 dt^2 + (N^\perp)^{-2}dr^2 + r^2 (\frac{4 G J}{r^2} dt + d\phi)^2,
\end{equation}
where $\phi$ is an angle and 
\begin{equation}
(N^\perp)^2 = -8 G M + \frac{r^2}{  \ell^2} + \frac{16 G^2 J^2}{r^2} = \frac{(r^2 - r_+^2)(r^2-r_-^2)}{r^2  \ell^2}
\end{equation}
the black hole horizons being given by 
\begin{equation}
r_\pm =  \sqrt{2 G \ell(\ell M + J)} \pm  \sqrt{2 G \ell(\ell M - J)}.
\end{equation}
The thermodynamic quantities can be computed to be (the superscript "+" refers to the outer horizon):
\bea\label{BTZParam}
M&=& \frac{r_-^2 + r_+^2}{8 G \ell^2}, \; J = \frac{r_- r_+}{4 G \ell}, \; T_H^+ =\frac{r_+^2-r_-^2}{2 \pi r_+ \ell^2},\\\Omega^+&=&\frac{r_-}{\ell r_+}, \; S^+ = \frac{\pi r_+}{2 G},
\eea
satisfying the first law
\be 
dM = T_H dS^+ - \Omega^+ dJ.
\ee
The entropy can be written as
\be \label{OuterSEH}
  S^+ = \frac{\pi^2 \ell}{3} (c_R T_R + c_L T_L),
\ee
where $c_L=c_R = \frac{3 \ell}{2 G}$ are the Brown-Henneaux central charges in pure GR and 
\be
T_{R/L} = \frac{r_+ \pm r_-}{2 \pi \ell^2}
\ee
 are related to the periods of the identification Killing vector leading to the BTZ black holes from $AdS_3$ and are interpreted as the right and left moving temperatures of the dual CFT \cite{Maldacena:1998bw}.
At the inner horizon, we have
\be\label{InnerQ}
T_H^- =\frac{r_+^2-r_-^2}{2 \pi r_- \ell^2},\; \Omega^-=\frac{r_+}{\ell r_-}, \; S^- = \frac{\pi r_-}{2 G},
\ee
satisfying
\be
  -dM  = T_H^- dS^- + \Omega^- dJ
\ee
in agreement with \cite{Castro:2012av}. The inner horizon is expressed in terms of the left and right-moving temperatures as
\be\label{InnerSEH}
  S^- = \frac{\pi^2 \ell}{3} (c_R T_R - c_L T_L).
\ee 
The Hawking tempatures of the inner and outer horizons moreover satisfy
\be
  \frac{1}{T_H^{\pm}} = \frac{1}{2} \left(\frac{1}{T_L} \pm \frac{1}{T_R}  \right).
\ee

\section{Inner Mechanics in TMG}
Topologically massive gravity (TMG) \cite{Deser:1981wh} is described by the following action: 
\begin{equation}
 I_{TMG} = \frac{1}{16 \pi G} \left[\int_M  d^3x \, \sqrt{-g} (R - 2 \Lambda) + \frac{1}{\mu} \; I_{CS} \right].
\end{equation}
The gravitational Chern-Simons term $I_{CS}$ is given by
\begin{equation} 
 I_{CS} = \frac{1}{32 \pi G} \int_M d^3x \, \sqrt{-g} \varepsilon^{\lambda \mu \nu} \Gamma^\alpha_{\lambda \sigma} \left(\partial_\mu \Gamma^\sigma_{\alpha \nu} + \frac{2}{3} \Gamma^{\sigma}_{\mu \tau}\Gamma^{\tau}_{\nu \alpha}\right),
 \end{equation}
where $G$ is the Newton's constant, $\Lambda = \pm1/\ell^2$ the cosmological constant with $\ell$ the $(A)dS_3$ curvature and $\mu$ is the Chern-Simons coupling which can be taken positive without loss of generality. 

%
The equations of motion for TMG with a cosmological constant are given by
\be
R_{\mu\nu} - \frac{1}{2}R g_{\mu\nu} - \frac{1}{\ell^2} g_{\mu\nu} + \frac{1}{\mu}C_{\mu\nu} = 0 \label{eomTMG}
\ee
where
\be
C_{\mu \nu} = {{\varepsilon_{\mu}}^{\alpha \beta}}\nabla_{\alpha}(R_{\beta\nu} - \frac{1}{4}g_{\beta\nu}R)
\ee
is the Cotton tensor and $\varepsilon^{\mu\nu\rho}$ is the Levi-Civita tensor. The above theory is known to have a rich set of vacua \cite{Nutku:1993eb,Gurses:2008wu,Anninos:2008fx,Chow:2009km,Anninos:2009jt} which have appeared in various recent applications of the gauge/gravity duality such as Kerr/CFT \cite{Guica:2008mu} and Holographic Condensed Matter \cite{Son:2008ye, Balasubramanian:2008dm}. Because the theory is higher-derivative, the Bekenstein-Hawking entropy is no longer given by the area of the horizon, but can be computed for any general-covariant using Wald's formula\cite{Wald:1993nt}. For TMG, the corrected entropy was computed in \cite{Bouchareb:2007yx, Solodukhin:2005ah, Tachikawa:2006sz,Kraus:2005zm,Park:1998qk}, taking into account the fact that the Lagrangian is diff-invariant only up to a boundary term. Writing the black hole metric in ADM form:
\be
ds^2 = -N(r)^2 dt^2 + \frac{dr^2}{M(r)^2} + R(r)^2 (d\theta^2 + N^\phi(r) dt)^2
\ee   
the TMG entropy reads
\be
  S_{TMG} = \frac{\pi R}{2 G} + \frac{\pi M R^2 (N^\phi)' }{4 G \mu N}
\ee
evaluated at the Cauchy horizon ($r_+$ or $r_-$).

The conserved charges associated with exact Killing vectors $\xi$ of the black hole solutions will also be modified with respect to their expressions in pure GR, and are given by 
\begin{equation}
\delta Q_\xi[g ;\bar g] =  \int_S \sqrt{-g}\, k^{\mu\nu}_\xi[\delta g ; g] \epsilon_{\mu\nu\rho}\,dx^\rho
\end{equation}
where the explicit form of the 1-form $ k^{\mu\nu}_\xi$ can be found in eq. (7) of \cite{Compere:2008cv}. In this expression, $\delta g$ typically is of the form $\delta g = (\frac{dg_{\mu \nu}}{dr_+} dr_+ + \frac{dg_{\mu \nu}}{dr_-} dr_- )dx^\mu dx^\nu$, where $r_\pm$ are the horizons of the black hole. Then $\delta Q_\xi$ simply computes the infinitesimal charge difference between solutions with parameters $(r_+,r_-)$ and $(r_+ + dr_+,r_- + dr_-)$, i.e. $\delta Q_{\p_t} \sim dM$ and $\delta Q_{\p_\phi} \sim dJ$.
All charges in the next sections are computed using this definition and \cite{CodeGeo}.
\newline
\subsection{BTZ black holes}
For $\Lambda < 0$, TMG admits BTZ black hole solutions with entropy
\be
   S_+ = \frac{\pi r_+}{2 G} + \frac{\pi r_-}{2 G \mu \ell},
\ee
i.e. it depends on the inner horizon. It can still be written in the Cardy form \re{InnerSEH} 
in which the Brown-Henneaux central charges for TMG are \cite{Park:2006gt, Sahoo:2006vz}
\be
   c_{L/R} = \frac{3 \ell}{2 G} (1 \mp \frac{1}{\mu \ell}).
\ee
The entropy associated with the inner horizon can similarly be computed to be
\be
   S_- = \frac{\pi r_-}{2 G} + \frac{\pi r_+}{2 G \mu \ell},
\ee
which also satisfies \re{InnerSEH}.

The inner Hawking temperature and angular velocities are the same as in \re{InnerQ}, while the mass and angular momentum become:
\be
  M_T = \frac{r_-^2+r_+^2}{8 G \ell^2} + \frac{r_-r_+}{4G \mu \ell^3}, \; J_T = \frac{r_-r_+}{4 G \ell} + \frac{r_-^2+r_+^2}{8 G \mu \ell^2},
\ee
where the subscript "T" refers to the TMG charges and not the original BTZ parameters \re{BTZParam}.
It is then easy to check that
\be
  -dM_T  = T_H^- dS^- + \Omega^- dJ_T
\ee
still holds in TMG.

The product of the inner and outer entropies is given by
\be
  S_+ S_- = \frac{\pi^2 \ell}{4} \frac{(M_T + J_T \mu)}{ G \mu}
\ee
and therefore depends on the mass of the black hole. In pure gravity however ($\mu \ra \infty$), we find it only depends on the angular momentum.
This can be made more transparent in writing
\bea\label{ProdBTZTMG}
  S_+ S_- &=& \frac{\pi^2 \ell}{6} (c_R E_R - c_L E_L) \\
      &=&  \frac{\pi^2 \ell}{12} \left( (c_R+c_L) (E_R-E_L) + (c_R-c_L) (E_R + E_L)\right) \nonumber
\eea
where $E_{L/R}$ are the left and right-moving energies:
\be
    E_{L/R} = \frac{\pi^2 \ell}{6} c_{L/R} T_{L/R}^2
\ee
related to the ADM charges by
\be
   M_T = E_R + E_L, \quad J_T = \frac{1}{\ell} (E_R - E_L).
\ee
We see from \re{ProdBTZTMG} that the mass dependence of the product of the areas is rooted in the diffeomorphism anomaly $c_L - c_R$ of TMG.

\subsection{Spacelike Warped Black Holes}

The spacelike warped black holes have the following metric\cite{Anninos:2008fx} (we set $\ell = 1$):
\begin{eqnarray}\label{0807WBH}
ds^2 &=& d\hat{t}^2 + \frac{d\hat{r}^2}{( \nu^2+3)(\hat{r} - r_+)(\hat{r}-r_-)} \\ &+&(2 \nu \hat{r}-\sqrt{r_+ r_-( \nu^2+3)})dt \;d \theta \nonumber+\frac{\hat{r}}{4}\{ 3( \nu^2-1) \hat{r} \\  &+&( \nu^2+3)(r_+ +r_-)-4 \nu \sqrt{r_+ r_-( \nu^2+3)}\}d\theta^2 \nonumber
\end{eqnarray}

Their ADM charges are (there is a sign difference in $J$ w.r.t. to \cite{Anninos:2008fx})
\bea
 M&=& Q_{\p_t} = \frac{(\nu^2 + 3)}{24 G \nu} \left(\nu(r_-+r_+) - \sqrt{r_- r_+ (3+\nu^2)}  \right) \\
 J&=& -Q_{\p_\theta} =   -\frac{\nu(\nu^2 + 3)}{96 G}[(\nuo)^2
 \nonumber\\
 &-& \frac{(5\nu^2+3)}{4\nu^2}({r}_+-{r}_-)^2]
 \eea
The Hawking temperature, angular velocity and entropy at the outer horizon were computed in \cite{Anninos:2008fx} and shown to satisfy the First Law.
TMG with $\Lambda < 0$ and  spacelike warped boundary conditions  was conjectured to be dual to a 2d CFT with central charges ($\nu := -\frac{\mu}{3}$, we'll consider $\nu > 0$ without loss of generality)
\begin{equation}
  c_R = \frac{5 \nu^2+3}{\nu G(\nu^2 +3)}, \quad  c_L = \frac{4 \nu}{(\nu^2 +3)G}.
\end{equation}
Spacelike warped black holes are obtained as quotient of global spacelike $WAdS_3$, allowing one to identify left and right-moving temperatures as 
 \begin{eqnarray}\label{TLR}
   T_R &:=& \frac{(\nu^2 +3)(r_+-r_-)}{8 \pi} \\
   T_L &:=& \frac{(\nu^2 +3)}{8 \pi} \left(r_+ + r_- -\frac{\sqrt{(\nu^2+3) r_+ r_-}}{\nu} \right )
 \end{eqnarray}
The Bekenstein-Hawking entropy associated with the outer horizon can be written as
 \begin{equation}
   S^+ = \frac{\pi^2}{3} (c_L T_L + c_R T_R). 
  \end{equation}
At the Inner Horizon, the Hawking temperature and angular velocity are
 \bea
   T_H^- &=& \frac{(r_+-r_- )(3+\nu^2)}{4 \pi (2 r_- \nu -\sqrt{r_+r_- (3+\nu^2)})}, \;\\ \Omega^- &=& \frac{2}{ (2 r_+\nu - \sqrt{r_+r_-(3+\nu^2)}}
 \eea
while the entropy can be computed to be
\be
  S^- = \frac{\pi}{24 G \nu} (-r_+ (3+\nu^2) -4 \nu \sqrt{r_-r_+(3+\nu^2)} + 3 r_-(1+3 \nu^2)),
\ee
and written as
\begin{equation}
   S^- = \frac{\pi^2}{3} (c_L T_L - c_R T_R). 
  \end{equation}
 Note that $S_-$ is obtained by simply  exchanging $r_+$ and $r_-$ in $S^+$.
The Inner First Law 
\be
-dM  = T_H^- dS^- + \Omega^- dJ
\ee
 is again satisfied.


The product $S_+ S_-$ is given by 
\be
   S_+ S_- = \frac{\pi^2 \ell}{6} (c_R E_R - c_L E_L),
\ee
where the left and right-moving energies are defined from the left and right-moving temperatures and given by \cite{Anninos:2008fx}
\bea
   E_L = \frac{3}{2} c_L G^2 M^2, \quad E_R = J +  \frac{3}{2} c_L G^2 M^2.
\eea
This can be rewritten is a slightly different way as
\be\label{ProductWarped}
    S_+ S_- = \frac{\pi^2 \ell}{6} \left(c_R (E_R - E_L) + (c_R - c_L) E_L \right).
\ee
We again see that the mass dependence of the product of the areas originates from the diffeomorphism anomaly.

\subsection{Warped de Sitter Black Holes}

Warped de Sitter black holes were first obtained in TMG with $\Lambda > 0$ in \cite{Nutku:1993eb,Gurses:2008wu}, and subsequently discussed in \cite{Bouchareb:2007yx,Anninos:2009jt}. Their holographic interpretation was discussed in \cite{Anninos:2011vd}. The metric reads
\bea
&&\frac{ds^2}{\ell^2} =  \frac{2\nu}{(3-\nu^2)^2} \left( 4\nu  \tilde{r} + \frac{ 3(\nu^2+1)\omega}{\nu {{}} } \right) d\tilde t d\tilde \theta \\  &+& \frac{d\tilde{r}^2}{(3 - \nu^2)(r_h - \tilde{r})(\tilde{r}+r_h)} + \frac{4\nu^2}{(3-\nu^2)^2}d\tilde t^2  \nonumber\\&+& \frac{ 3(\nu^2 + 1)}{ {{(3-\nu^2)^2}}} \left( \tilde{r}^2 + 2 \tilde{r} \omega  + \frac{(\nu^2-3)r_h^2}{3(\nu^2+1)}  + \frac{3(\nu^2 + 1)\omega^2}{4\nu^2} \right)d\tilde \theta^2\nonumber. \label{wdsbc}
\eea
The TMG parameter $\nu$ is restricted to the range $\nu^2 < 3$ (for $\nu^2 >3$, we recover the $WAdS_3$ black holes \cite{Anninos:2009jt}). An interesting feature of these spaces is that they exhibit both a black hole and a cosmological horizon located at $-r_h$ and $r_h$ respectively.
The corresponding left and right temperatures read \cite{Anninos:2009jt,Anninos:2011vd}
\be\label{TLRWdS}
T_L = \frac{3(1 + \nu^2)\omega}{8\pi\ell \nu^2}~, \quad T_R =  \frac{r_h}{2\pi\ell}~.
\ee
As already noted in \cite{Anninos:2011vd}, the entropies of the cosmological and black hole horizons can be written as
\be
S_{c} = \frac{\pi^2 \ell}{3}\left( c_L T_L + c_R T_R \right), \quad
S_{BH} = \frac{\pi^2 \ell}{3}\left( c_L T_L - c_R T_R \right)
\ee
with 
\be
c_L = \frac{4\nu \ell}{(3-\nu^2)}~, \quad c_R = \frac{(5\nu^2 - 3)\ell}{\nu(3 - \nu^2)}~,
\ee
in agreement with the previous observations.

The first law of thermodynamics was shown to be satisfied for the cosmological horizon,
\be
\delta Q_{\partial_{\tilde{t}}} = {T}_H^c \delta S_{c} + {\Omega}_c \delta Q_{\partial_{\tilde \theta}}~,
\ee
and one can check that a similar relation holds for the black hole horizon:
\be
-\delta Q_{\partial_{\tilde{t}}} = {T}_H^{BH} \delta S_{BH} - {\Omega}_{BH} \delta Q_{\partial_{\tilde \theta}}~,
\ee
where the charges, Hawking temperature and angular velocity at the cosmological horizon have been computed in \cite{Anninos:2011vd}. The intensive variables at the black hole horizon are 
\bea
 {T}_H^{BH}  &=&  \left|\frac{2 r_h \nu^2}{\pi(3(1+\nu^2)\omega)-4 r_h \nu^2}\right |,\\
 {\Omega}_{BH}  &=& \frac{4 \nu^2}{4 r_h \nu^2 - 3(1+\nu^2)\omega}.
\eea
As in \re{ProductWarped}, the product $S_c S_{BH}$ depends on $E_L$ because $c_L \ne c_R$, hence depends on the mass since $E_L \sim \delta Q_{\partial_{\tilde{t}}} $ and $E_L - E_R \sim \delta Q_{\partial_{\tilde{\phi}}}$ as for the Warped $AdS_3$ black holes.


\begin{center}
\section{Comment on New Massive Gravity}
\end{center}
Another unitary higher-derivative 3d gravity theory has attracted much attention recently, extending the Einstein-Hilbert term by a parity-preserving fourth-order term. Its action is
\be
  S = \frac{1}{16 \pi G} \int d^3 x \; \sqrt{-g} \left[R - \frac{1}{m^2} K- 2 \Lambda \right],
\ee
$K = R_{\mu \nu} R^{\mu \nu} - \frac{3}{8} R^2$. This theory admits BTZ as well as Warped black holes as solutions. The thermodynamic quantities have been computed in \cite{Clement:2009gq}. For BTZ, the mass, angular momentum and entropies are renormalized by the same factor (depending on $m$ and $\Lambda$) with respect to their pure GR expressions. Since $T_\pm$ and $\Omega_\pm$ are unchanged, the Inner First law and the fact $S_+ S_-$ depends only on $J$ trivially holds. This can actually be extended to a general diffeomorphism-invariant theory of gravity. It was shown in \cite{Saida:1999ec} that for a general diff-invariant Lagrangian of the form $f(R_{\mu \nu},g_{\mu \nu})$, the entropy computed using Wald formula \cite{Wald:1993nt} only gets renormalized by a factor $g^{\mu \nu} \frac{\p f}{\p R_{\mu \nu}}$ with respect to its pure gravity expression (note that this is not true in TMG). On the other hand, the first law associated with the outer horizon is supposed to hold for general $f$\cite{Iyer:1994ys, Iyer:1995kg}. This means that $M$ and $J$ must be renormalized accordingly, hence both statements -- Inner First law being satisfied and $S_+ S_-$ independent on $M$-- generalize. This can also be noticed by remarking that the outer and inner horizon entropies are still given by \re{OuterSEH}, \re{InnerSEH} with \cite{Kraus:2006wn}
\be
   c_L = c_R = \frac{\ell}{2 G} g^{\mu \nu} \frac{\p f}{\p R_{\mu \nu}},
\ee
while the dependence of $T_{L/R}$ and $E_{L/R}$ on the ADM charges $M$ and $J$ is the same as in pure GR up to a factor \cite{Saida:1999ec}.
For Warped $AdS_3$ black holes, one cannot connect to the pure gravity computation, since these are not Einstein backgrounds, nor are the central extensions known for a general higher-derivative theory (Warped black holes don't have constant curvature). For New Massive Gravity however, using the explicit expressions of the charges computed in \cite{Clement:2009gq}, one finds that the inner first law holds and that $S_+ S_-$ is independent of $M$, in agreement with \re{ProductWarped}.
\newline
\newline
\section*{Acknowledgments}


I would like to thank T. Azeyanagi, S. de Buyl, A. Castro, Griffus, T. Jacobson, G. Ng, and M.-J. Rodriguez for very useful discussions, A. Castro and M.-J. Rodriguez for sharing their results prior to publication and A. Castro for her careful reading of the manuscript and helpful suggestions. I would also like to thank the organizers of the Program "Bits, branes and black holes" and the KITP, where part of this work was done, for hospitality. I acknowledge support from the Fundamental Laws Initiative, of the Center for the Fundamental Laws of Nature, Harvard University.

\end{document}